\documentstyle[draft,floats,prd,amstex,amssymb,psfig,aps,prd]{revtex}

\begin{document}

\author{ Benjamin D.\ Wandelt$^{1,2}$, Eric Hivon$^{1,3}$ and
Krzysztof M.\ G\'orski$^{1,4,5}$}
\address{$^1$Theoretical Astrophysics Center,
        Juliane Maries Vej 31,
        DK-2100 Copenhagen \O,
        Denmark}
\address{
        $^2$Department of Physics,
        Princeton University,
        Princeton, NJ 08540,
        USA
        }
\address{
	$^3$Observational Cosmology,
	Caltech,  Mail Code 59-33,
	Pasadena, CA 91125,
	USA
        }
\address{
	$^4$European Southern Observatory, Garching bei M\"unchen, Germany
        }
\address{
        $^5$Warsaw University Observatory, Warsaw, Poland
        }

\title{The Pseudo-$C_l$ method: cosmic microwave background anisotropy power spectrum statistics for high precision cosmology}

\maketitle
\begin{abstract}
As the era of high precision cosmology approaches,
the empirically determined
power spectrum of the microwave background anisotropy
$C_l$ will provide a crucial test for cosmological theories. 
We present an exact semi--analytic framework 
for the study of the ampling statistics of the $C_l$ resulting
from  observations with partial sky coverage
and anisotropic noise distributions. This includes space--borne, air--borne
and ground--based experiments.
We apply this theory  to  demonstrate its power for constructing fast but
unbiased approximate methods for the joint estimation of cosmological
parameters. 
Further applications, such as a test for possible non--Gaussianity of
the underlying theory and a ``poor man's power spectrum estimator''
are suggested. An appendix derives recursion relations for the
efficient computation of the couplings between spherical harmonics
on the cut sky.
\end{abstract}
\date{\today}
\pacs{PACS Numbers: 98.70.Vc,98.80.-k,95.75.Pq,02.50.-r}

\section{Introduction} 
\label{quad:intro}
During the next few years ground based observations and
balloon missions \cite{nonsatellite} as well as
satellite observation \cite{map,planck} promise exquisite
determinations of the angular distribution of 
cosmic microwave background (CMB) anisotropies. 

Inflationary cosmogonies, 
the most popular theories of structure formation,
predict a statistically
isotropic CMB with Gaussian fluctuations.
To linear order in perturbation theory this means that within this
class of models all cosmologically relevant information contained in
the map of anisotropies 
is distilled in the angular power spectrum  $C_l$.

An accurate estimate of these quantities therefore promises to have
tremendous impact
on our knowledge of the global properties of the universe and
the physical processes which lead to the formation of
structure\cite{Knox95,jungman,ZSS,BET}. 
This knowledge is quantified in terms of 
cosmological parameters such as  $\Omega_{tot}$, $\Omega_b$, $H_0$, {\em etc}.

The $C_l$  are uniquely defined as rotationally invariant quadratic combinations
of the coefficients of an
expansion of the anisotropy $\frac{\Delta T}{T}$ in
spherical harmonics. Their useful statistical properties rest on the 
the fact that the spherical harmonics are a complete and
orthogonal basis set on the sphere. However, in realistic experiments,
only part of the celestial sphere is covered, if only
because our CMB sky is obscured by the 
Milky Way. Also, due the particulars of the
observational strategy,
non--uniform and possibly correlated noise 
contaminates the observed signal. 
In order to interpret past observations and forecast the impact of
future observations we need a framework for studying the
sampling statistics of the $C_l$ for incomplete sky coverage  in
the presence of  non--uniform noise.

\begin{figure}[tf]
\centerline{\psfig{file=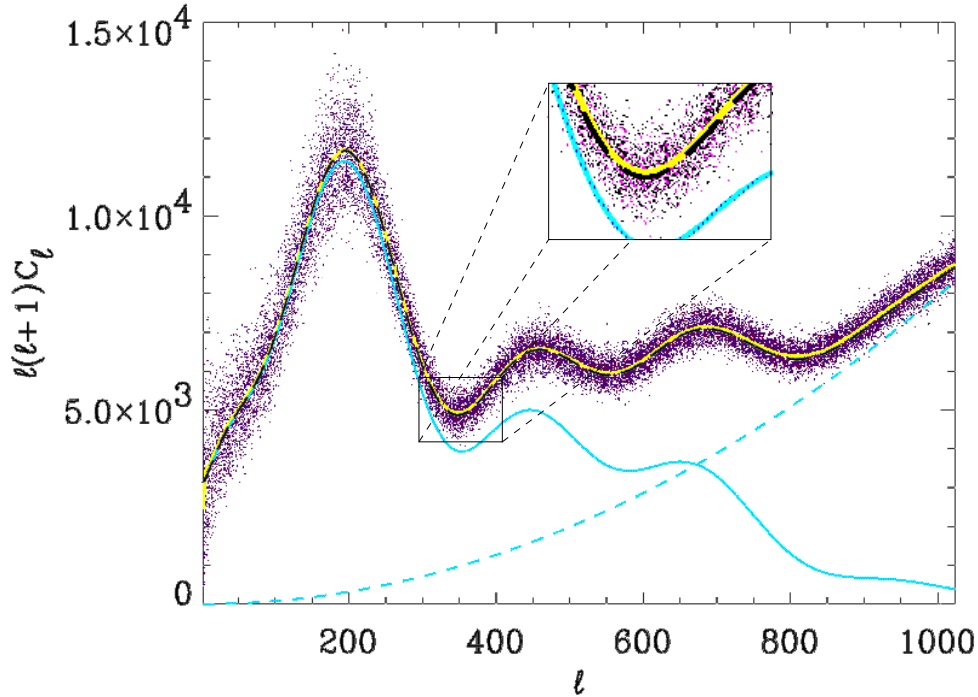,width=.8\textwidth}}
\caption[Pseudo--$C_l$ Monte Carlo simulations.] 
{The dots are 20 of our 3328 Monte Carlo
 simulations. On these we superpose the Monte Carlo average (yellow) and
the theoretical $C_l$ (black dashed),
which are the sum of signal (blue, solid) and noise (blue, dashed).
The discrepancy seen at low $l$, and to a lesser degree in the inset,
demonstrates the effect of correlations due to partial sky coverage
even for a satellite experiment capturing 70\% of the CMB sky.}
\label{fig:mc}
\end{figure}

It is worth noting that there are several different kinds of $C_l$
which we will be referring to in this paper. 

The
stochastic nature of the observed $C_l$ is illustrated in Figure \ref{fig:mc}, where the
straightforward computation of $C_l$ from a subset of our Monte Carlo simulations
of incomplete and noisy skies are overlaid on the theoretical predictions.
One or more 
of the following simplifying assumptions can be invoked 
in  cases where one would like to avoid the full complexity of the
problem ({\em e.g.\ } \cite{Knox95,jungman,lineweaver1,lineweaver2}):
\begin{itemize}
\item (Independence) The
observed $C_l$ are approximately independent due to nearly full and
uniform sky
coverage, 
\item (Scaled $\chi^2$) Their sampling distributions do not change appreciably
from the  $\chi^2$ distributions which apply for the full sky, apart
from a rescaling of the number of degrees of freedom by the sky
fraction $f$, and

\item (Gaussianity) Their sampling distributions are well--approximated
by Gaussian distributions which have the same first and second moments as these
rescaled $\chi^2$ distributions. 
\end{itemize}

There are 
several  reasons to carefully assess these approximations and, if
necessary, go beyond them. 
One reason is that in the short term 
balloon and ground based experiments
will provide the leading edge science results to the field. Due to
practical limitations we cannot hope to even come close
to full sky coverage with these types of experiments. The observation
regions are often ring--shaped or cover a circular region of the sky
which subtends a small solid angle.  It is an urgent matter to assess
statistically how these imminent experiments can constrain the power
spectrum and cosmological parameters.

Further, the planned satellite missions are aiming to
determine the $C_l$ to sub--percentage accuracy. 
In the case of the Planck Surveyor mission,
this has given us the hope
of detecting small effects such as 
secondary anisotropies which are due to nonlinear
gravitational effects on the CMB photons during the free--streaming
epoch. An example of such an effect is anisotropies due to
gravitational lensing. This 
is an important issue because a detection would break otherwise
present parameter
degeneracies and allow a consistent parameter estimation from CMB data
alone \cite{SE}.  

More generally, 
the impact CMB observations will have on cosmology makes it important
to use  approximations 
in a controlled way. For example, in the
analysis of COBE--DMR data it was realized that 
using Gaussian approximations for
quadratic quantities introduced systematic biases \cite{banday}.
At the same time, we need approximations to make feasible the analysis of
the huge CMB data sets we expect in the coming years.

The Maximum Likelihood Inversion (MLI) 
method ({cf.\ } \cite{gorski0,gorski1,gorskimoriond,BJK} for
applications in a cosmological context) is the standard framework for solving
this type of problem.
To illustrate, we write down the likelihood functional for a Gaussian
sky. In order to do so we need to define the vector of observed pixel
temperatures on the sky ${\bf {d}}$ with $N_{pix}$ components. 
A given theory and observational strategy on the sky  imply
the covariance structure of the elements $d_i$ in terms of 
signal and noise covariance matrices
${\bf {S}}$ and ${\bf {N}}$ such that
\begin{equation}
\left\langle {d_i d_j} \right\rangle=S_{ij}+N_{ij}.
\end{equation}
The signal covariance matrix $S$ is simply a discretised version of
the theoretical angular correlation function 
\begin{equation}
\xi\!\left(\alpha_{ij}\right)=\sum_{l=2}^{\infty}\frac{2l+1}{4\pi}C_l P_l(\alpha_{ij}),
\end{equation}
evaluated at the pixel locations. The
noise covariance ${\bf {N}}$ reflects those imperfections of the
experiment and observational strategy which lead to spatial
correlations in the projected noise.

This leads to an expression for the  likelihood
\begin{equation}
L(d | \mathrm{theory})=
\frac{\exp{\left(-{\frac 12}
{\bf {d^T}}({\bf {S}}+{\bf {N}})^{-1}{\bf {d}}\right)}}
{\sqrt{(2\pi)^{N_{pix}}\left\vert {{\bf {S}}+{\bf {N}}}\right\vert}}
\label{likelihood}
\end{equation}
The matrix ${\bf {S}}$ is a  function of 
predicted $C_l$, which are in turn determined by   the
parameters of some theory. The matrix ${\bf {N}}$ includes a model of
the expected noise covariance given the experimental setup.
The structures of both  ${\bf {S}}$ and ${\bf {N}}$  also depend on
the observed region on the
sky. 

For a given set of data the likelihood is then considered as a
function of the parameters of a given theory.
The maximum likelihood parameter estimates are
determined as those which maximise  Eq.\ (\ref{likelihood}).
While these manipulations are conceptually simple, the practical
difficulties are apparent. The number of operations
needed to evaluate  Eq.\ (\ref{likelihood}) (or equivalently an estimator 
derived from it) scales as the most costly operations contained in it.
These  are the determinant evaluation in the denominator and the matrix
inverse in the exponent. For a general situation both of these require $N_{pix}^3$
operations. 

In the near future
$N_{pix}$ will be $10^5$ from balloon experiments (such as
Maxima), then $10^6$ for MAP and in less than a decade 
$10^7$--$10^8$ for Planck.
These numbers imply $10^{15}$--$10^{24}$
computations. 
Even when considering future growth in computing
performance, estimates of the computational time required to evaluate
 Eq.\ (\ref{likelihood}) a single time reach up to many years
\cite{borrill}. To compute the maximum likelihood estimator requires
many likelihood evaluations and would thus require an unfeasibly long
period of time. 
For practical purposes, straightforward
applications of the MLI method therefore fail to give answers in
finite computational time.

Previous studies of this subject have usually directly addressed the
problem of "solving for the $C_l$". Tegmark \cite{tegmark} rejects the
maximum likelihood estimator on the grounds that it is too time--consuming
to compute and suggests a
minimum variance (and hence ``optimal'') method based on a quadratic
estimator which 
makes a form of the Gaussianity assumption above. While this method is
powerful in the regime where it is applicable, 
it does not  provide a 
means of assessing the validity of this assumption. Within our analysis
we can make the connection between  the minimum variance and the maximum
likelihood methods ({cf.\ } subsection \ref{leastsquaresest}). 

The approach of Bond, Jaffe and Knox \cite{BondJaffeKnox2} is again very
different from ours. It consists in finding  an approximate form for the likelihood
and parametrising the shape of the power spectrum in terms of a set of
top hat functions, whose height gives the average value of the $C_l$
over its width. The width of these top hats is chosen wide enough that
correlations between them become small. Their method is of practical
significance because their approximate ansatz for the likelihood is
easy to evaluate.  

Our perspective is the following: we endeavour to provide an {\em ab
initio} formalism which lays bare the statistical properties of the
power spectrum coefficients. It then turns out that the insight gained
from this study can be used to formulate unbiased and well--controlled methods for
parameter estimation.

\begin{figure}[tf]
\centerline{\psfig{file=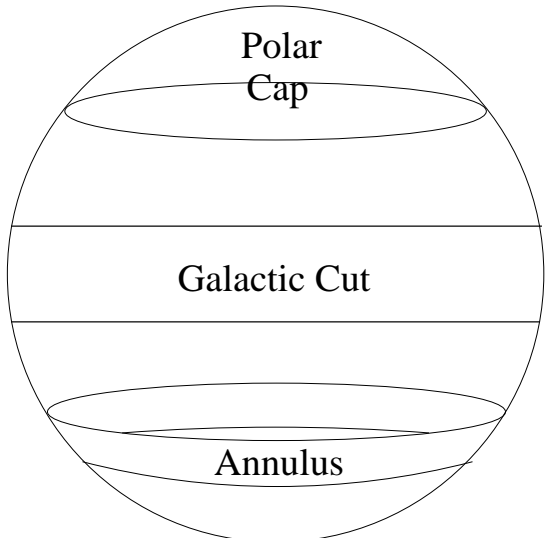,width=.4\textwidth}}
\caption[Sky observation geometries for which the formalism is exact.] {Annuli,
circular patches such as polar caps and even disjoint geometries such as
the full sky with an annulus (such as a Galactic cut) removed are
within the class for which our formalism is exact.  }
\label{fig:cutsky}
\end{figure}

To provide a quantitative basis for this discussion  we
present in this paper an efficient 
semi--analytic calculational framework within which the
statistical properties of the $C_l$ power spectrum coefficients can be
studied for theories which generate a Gaussian CMB sky.
This framework is exact for the important class of survey geometries and noise patterns
which obey rotational symmetry about one 
arbitrary axis and is not limited to large sky coverage.
This class contains both skies with symmetric and asymmetric cuts
north and south of the Galactic equator, rings, annuli, and polar cap shaped
regions of any size as indicated in Figure \ref{fig:cutsky}.
For this class we obtain 
exact analytical solutions for the mean $\left\langle {C_l} \right\rangle$, the variance
$\left\langle {\left(\Delta C_l\right)^2} \right\rangle$, the skewness, kurtosis and 
all higher moments of the sampling distribution for all $l$. 
We derive exact and conveniently computable 
analytic expressions for the marginalised probability distributions
of the cut sky $C_l$ for all $l$ as well as integral representations for
the joint distribution of all $C_l$. 
As an added bonus we find that our methods also allow dealing with arbitrary noise patterns
to very good accuracy and conclude by illustrating their use in some 
applications.

To test our results we perform state--of--the--art Monte Carlo simulations for a high
resolution CMB satellite with a Galactic cut of size 
$\pm 20^\circ$ (resulting in a
sky coverage comparable to COBE). We use a sine modulated noise
template as a simple 
model of the noise pattern which
would result from scanning along great circles through the
ecliptic poles. The numerical results we obtain provide a solid check on our
analytic calculations. 

The plan of this paper is as follows: 
in section \ref{quad:general} we lay out the general framework.
Section \ref{quad:pclstats} develops the statistical theory of power
spectrum coefficients for observation regions which are azimuthally
symmetric about some axis. We 
arrive at expressions for the cut--sky moments of any order and solve
the moment problem to obtain the exact sampling
distributions. We compare our
analytical results to Monte--Carlo simulations in section
\ref{quad:results}. Several applications of our methods are presented
in section \ref{quad:applications}. Section \ref{quad:conclusions} contains 
further discussion and our conclusions. Appendix \ref{quad:loworder}
discusses the effects of non-cosmological foregrounds on the results
presented in this paper. Details of our analytical and
numerical techniques are given in three further appendices.

\section{The framework} \label{quad:general}
The full sky of CMB temperature fluctuations can
be expanded in spherical harmonics, $Y_{lm}$,
as
\begin{equation}\label{quad:alms}
T({\bf \gamma})=\sum_{l=0}^{l_{max}}\sum_{m}a_{lm}Y_{lm}(\gamma)
\end{equation}
where ${\bf {\gamma}}$ denotes a unit vector pointing at polar angle $\theta$ and
azimuth $\phi$. Here we have assumed that there is insignificant signal power in modes
with $l>l_{max}$ and introduce the convenient notation that all sums over $m$ run from
$-l_{max}$ to $l_{max}$ but all quantities with index ${lm}$ vanish
for $m>l$.

In a universe which is globally isotropic and contains Gaussian fluctuations, the $a_{lm}$ are
independent, Gaussian
distributed with zero mean and specified variance 
$C^{theory}_l\equiv\langle\left\vert {a_{lm}}\right\vert ^2\rangle$. 
Hence, for noiseless, full sky measurements,
each measured $C_l$ independently follows a
$\chi^2$--distribution with $2l+1$ degrees of freedom and mean $C^{theory}$.
Owing to Galactic foregrounds, limited surveying time or other
constraints inherent in the experimental setup, the
temperature map that comes out of an actual measurement will be
incomplete. In addition, a given scanning strategy will produce a
noise template: high noise per pixel results in
regions where the scanning lines are less densely packed compared to
regions where they are denser. We model the noise as a Gaussian field
with zero mean which is independent from pixel to pixel and modulated
by a spatially varying root mean squared amplitude  $W_N({\bf {\gamma}})*W({\bf {\gamma}})$ (the factor of $W$ appearing
because the noise too is zero in unobserved regions).
 Therefore the
observed temperature 
anisotropy map is in fact
\begin{equation}
\tilde{T}({\bf \gamma})=W({{\bf {\gamma}}})\left[T({{\bf {\gamma}}})+W_N({{\bf {\gamma}}})T_N({{\bf {\gamma}}})\right]
\label{realtemp}
\end{equation}
where $W$ is unity in the observed region and zero elsewhere.

The coefficients in the spherical harmonic expansion 
recovered from such an incomplete observing region are
\begin{eqnarray}
\tilde{a}_{l'm'} & = & \int_{{\cal O}} d\Omega\: Y^{\ast}_{l'm'}({{\bf {\gamma}}}) T({{\bf {\gamma}}})
\newcommand\Tfluct{\frac{\delta T}{T}({\bf {\gamma}})}
\nonumber\\
         & = & \sum_{lm} {a_{lm}} \int_{{\cal O}} d\Omega\: Y^{\ast}_{l'm'}({\bf {\gamma}}) Y_{lm}({\bf {\gamma}})
\label{eq:palms}
\end{eqnarray}
The notation ``$\int_{\cal O}$'' denotes integration over the observed region.
Note that the usual orthogonality property of the $Y_{lm}({\bf {\gamma}})$ does not hold
any longer because we do not integrate over all solid angle. This
becomes clearer if we define the  {\em geometric coupling matrix} 

\begin{equation}
	W_{l'm'\;lm} \equiv \int_{{\cal O}} d\Omega\: Y^{\ast}_{l'm'}({\bf {\gamma}}) Y_{lm}({\bf {\gamma}})
\label{Wlmlmdef}
\end{equation}
and write 
\begin{equation}
\tilde{a}_{l'm'} =  \sum_{lm} W_{l'm'\;lm} {a_{lm}}.
\label{eq:palms1}
\end{equation}
By definition, the $W_{l'm'\;lm}$ are just the matrix elements of the 
$W({{\bf {\gamma}}})$ window in a spherical harmonic basis. These
matrices have been previously discussed in a similar context in the
pioneering papers by 
Peebles and Hauser \cite{Peebles,HauserPeebles}. 

This means that expanding $\tilde{T}$ 
as in  Eq.\ (\ref{eq:palms}) produces a set of {\em correlated} Gaussian
variates $\tilde{a}_{lm}$ for the signal and $\tilde{a}_{N\;lm}$ for the
noise. These combine into
power spectrum coefficients
\begin{equation}
\tilde{C}_l=\frac1{2l+1}\sum_m
\left\vert{\tilde{a}_{lm}+\tilde{a}_{N\;lm}}\right\vert ^2
\label{pcldef}
\end{equation}
whose statistical properties differ from the ones of the $C_l$. 
We therefore refer to these quantities as {\it pseudo}-$C_l$. 
In what follows we will discuss the statistical
properties of these quantities.

\section{Exact pseudo--$C_\l$ statistics}
\label{quad:pclstats}

Let us focus on the terms of the sum  Eq.\ (\ref{pcldef}). They are squares of
Gaussian distributed variates with zero mean. It follows that the
$\tilde{C}_l$ are sums of $2l+1$ $\chi^2$ variates with one degree of
freedom. However, each term in the sum has a  different expectation
value and may in general be correlated, so the $\tilde{C}_l$ are {\em
not} $\chi^2$ distributed with 
2l+1 degrees of freedom. 

If we assume white noise, $C^N_l\equiv C^N$,
then each term in the sum  has expectation value 
\begin{equation}
\sigma^2_{lm}=\frac{\sum_{l'm'}C_{l'}\left\vert{W_{l'm'\;lm}}\right\vert
	^2+
	C^N\!\!\int_O d{{\bf {\gamma}}} W_N({{\bf {\gamma}}})^2\lambda^2_{lm}(\theta )}{2l+1}
\label{slm}
\end{equation} 
A few remarks are in order:
\begin{enumerate}
\item  Azimuthal symmetry has not been invoked to
derive this relation.
\item The cross--term which comes from expanding the square in  Eq.\ (\ref{pcldef})
has vanishing expectation value because signal and noise are assumed
to be uncorrelated.
\item The sum over $C_l$ includes the non--cosmological $l=0$ and $l=1$
terms. The appearance of these terms merits some further discussion,
which we relegate to Appendix \ref{quad:loworder}.
\end{enumerate}

\subsection{Digression: connecting to the least squares estimator}
\label{leastsquaresest}
Summing  Eq.\ (\ref{slm})  over $m$ gives a set of linear equations which also
results in a different context, namely  after   
minimising the variance of the $C_l$ estimator in Tegmark \cite{tegmark}.  There it appears as a set of linear equations for the estimator, which the author assumes to be Gaussian distributed. 

We see from our treatment that this equation is only valid  after the expectation values are taken. The quantities involved are {\em quadratic} in Gaussian variates which leads to the presence of a cross term which only disappears after taking the expectation value. 

This is the reason why some $C_l$ estimates in \cite{tegmark} turn out
negative for values of $l$ where the signal to noise ratio becomes of
order unity. This is of course unphysical because the $C_l$ are
properly quadratic and hence positive semi--definite quantities. We
observe that it is precisely neglecting this fact which leads to
unphysical results. 

On the other hand, our treatment provides alternative proof of the
fact that the least--squares estimator is unbiased. It also explains the
significance of the Fisher matrix which appears in this context: is simply a measure of the
geometric couplings between eigenmodes of total angular momentum. 

\subsection{Azimuthal symmetry}

To elucidate the correlation structure between the $\tilde{a}_{lm}$,
 Eq.\ (\ref{eq:palms1}), we
write  Eq.\ (\ref{pcldef}) as
a quadratic form 
\begin{equation}
\tilde{C}_l=\frac1{2l+1}\sum_{m}\:\sum_{l',m',l'',m''}\xi^\ast_{l'm'}{\cal M}^{(lm)}_{l'm'l''m''}\xi_{l''m''}
\label{quadform}
\end{equation}
in  independent normal variates $\xi_{lm}$ with zero mean and unit
variance. The object ${\cal M}$ describes the combined effect of
statistical
correlation  and the geometrical couplings between the
various expansion coefficients.

For an isotropic Gaussian field on the full sky
\begin{equation}
{\cal{M}}^{(lm)}_{l_1m_1\;l_2m_2}=
\delta_{ll_1}\delta_{mm_1}\left(C^{theory}_l+C^{noise}_l\right)\delta_{ll_2}\delta_{mm_2}.\quad\text{(isotropic,
full sky)}
\end{equation}

We show in Appendix \ref{quad:factorise}
that  for a Gaussian field with an anisotropic but azimuthally
symmetric weight function $W(\theta)$, ${\cal M}$ is of the form
\begin{equation}
{\cal{M}}^{(lm)}_{l_1m_1\;l_2m_2}=\delta_{mm_1}W^{(l)}_{l_1m_1}\left(C^{theory}_l+C^{noise}_l\right)W^{(l)}_{l_2m_2}\delta_{mm_2},\quad\text{(anisotropic,
cut sky)}\label{Mstruct}
\end{equation}
where
\begin{equation}
{\cal W}_{l'm}^{(l)}\equiv W_{l'm\;lm} =\int 
d\mu W(\mu) \lambda_{lm}(\mu)\lambda_{l'm}(\mu),
\label{eq:Mazim}
\end{equation}
and we  assume  identical  windows for  signal and 
noise for notational simplicity in  Eq.\ (\ref{Mstruct}). The generalisation
for arbitrary 
azimuthally symmetric noise windows is straightforward.


The structure exhibited in Eqs.~(\ref{Mstruct}) and  (\ref{eq:Mazim}) is the key fact which allows the derivation and cheap
evaluation of the exact results we obtain.
An efficient algorithm for the
evaluation of the matrix elements of $W$ is given in appendix
\ref{quad:couple}. For a
maximum $l$ of 1024 this takes just over 
one minute on a single R10000 CPU. 

It follows that while the $\tilde{a}_{lm}+\tilde{a}_{N\;lm}$ terms in
 Eq.\ (\ref{pcldef})
are correlated for
different $l$, they will be
{\em independent} for different $m$. As a consequence, under the
assumption of azimuthal symmetry the
$\tilde{C}_l$ are sums of independent 
Chi--squared ($\chi^2$) variates, each with one degree of freedom but
different expectations. Below we succeed in finding analytical
expressions for the probability density functions
of such  generalised $\chi^2$ variates and we hence 
solve for the statistical properties of the pseudo--$C_l$ exactly.  

\subsection{Pseudo--$C_{\l}$  moments}
\label{quad:moments}

The problem is to compute the statistical properties
of a sum of independent variates, each
of which has a known distribution. A useful tool for attacking this
kind of problem is the method of 
characteristic functions \cite{KendallStuart}. Appendix \ref{quad:cfs}
provides a heuristic introduction to this method.

The characteristic function of a
$\chi^2$ variate with one degree of 
freedom is given by 
\begin{equation}
\frac{1}{(1-2ix\sigma_{lm}^2)^{{\frac 12}}}.
\end{equation}
We can multiply them together for each term in the sum  Eq.\ (\ref{pcldef})
and inverse Fourier transform to obtain
$p(\tilde{C}_l)$ as
\begin{equation}
p(\tilde{C}_l)= \frac{1}{2\pi}\int_{-\infty}^{+\infty}dx 
\frac{e^{-i\tilde{C}_l x}}{\prod_{m=-l}^{l}(1-2ix\sigma_{lm}^2)^{{\frac 12}}}
\label{charfunc}
\end{equation}
The log of the integrand in this expression (not including the complex
exponential) is the cumulant generating
function, which we can differentiate to obtain {\em all} cumulants of the
pseudo--$C_l$ as 
\begin{equation}
\kappa_n=2^{n-1}(n-1)!\sum_m(\sigma_{lm}^2)^{n}.
\end{equation}

We give the following expressions
for the mean, variance, skewness 
$\beta_1=\frac{\langle(\Delta\tilde{C}_l)^3\rangle}{\langle(\Delta\tilde{C}_l)^2\rangle^{\frac32}}$
and kurtosis
$\beta_2=\frac{\langle(\Delta\tilde{C}_l)^4\rangle}{\langle(\Delta\tilde{C}_l)^2\rangle^2}$ as examples:
\begin{equation}
\begin{tabular}{cc}
$\langle\tilde{C}_l\rangle=\sum_m \sigma^2_{lm}$,& 
   $\langle(\Delta\tilde{C}_l)^2\rangle=2\sum_m\sigma^4_{lm}$\\
\\
$\beta_1=2^{\frac32}\frac{\sum_m{\sigma_{lm}^6}}{\left(\sum_m{\sigma_{lm}^4}\right)^{\frac32}},$&
$\beta_2={12}\frac{\sum_m{\sigma_{lm}^8}}{\left(\sum_m{\sigma_{lm}^4}\right)^2}$.\\
\end{tabular}
\label{meanetc}
\end{equation}

\subsection{Pseudo--$C_{\l}$  distributions}
\label{quad:probdist}
It turns out that we do not have to contend with just the cumulants. 
We will now derive the analytical form for the pseudo--$C_l$
distributions.
We define $s^{(l)}_m=(2/\sigma^2_{lm})$ and note that 
$s^{(l)}_m =s^{(l)}_{-m}$. Then all terms with equal $\left\vert{m}\right\vert\geq 1$ pair up
and we obtain
\begin{equation}
p(\tilde{C}_l)= \frac{A^{(l)}}{2\pi}\int_{-\infty}^{+\infty}dx 
\frac{e^{-i\tilde{C}_l x}}{(s^{(l)}_0-ix)^{{\frac 12}}\prod_{m=1}^{l}{(s^{(l)}_m-ix)}}
\end{equation}
where $A^{(l)}\equiv\sqrt{s^{(l)}_0}\prod_{m=1}^{l}s^{(l)}_m$. We
expand in partial fractions and integrate each term analytically (using a
standard integral from \cite{GradshteynRyzhik}) to
produce a closed form solution in terms of 
incomplete gamma functions $\gamma(\alpha ,x)$:
\begin{equation}
\!\!\!\!\!\!\!\!p(\tilde{C}_l)={A'}^{(l)}
\sum_{m=1}^{l}\frac{\exp\left(-s^{(l)}_m \tilde{C}_l\right)\,\gamma\!\left(\frac{1}{2},(s^{(l)}_0-s^{(l)}_m)\tilde{C}_l\right)}
{\sqrt{(s^{(l)}_0-s^{(l)}_m)}
\prod_{\;m=1}^{'\;l} (s^{(l)}_{m'}-s^{(l)}_{m})}
\label{probdist}
\end{equation}
where the primed product symbol $\prod^{'}$ only multiplies
factors 
which have $m\neq m'$ and the normalization constant is
\begin{equation}
{A'}^{(l)}\equiv\frac{\sqrt{s^{(l)}_0}\prod_{m=1}^{l}s^{(l)}_m}{2
\Gamma(\frac{3}{2})}. 
\end{equation}

We have, therefore, reduced the problem of determining the sample
probability density of the pseudo--$C_l$ to computing the
$\sigma_{lm}^2$. The computation of these coefficients has to be done
only once for a given survey geometry and model theory. In fact it is
clear from their definition and  Eq.\ (\ref{slm}) that once the ${\cal W}^{(l)}_{l'm}$ are
computed for a given survey geometry, the $\sigma_{lm}^2$ can be
generated very easily given a model $C_l$ spectrum.

\subsection{Correlations between different $\tilde{C}_{\l}$}
\label{joint}

The distributions we have just derived are 1-point or marginal
distributions. They contain complete statistical information about each
$\tilde{C}_l$ taken for itself, but they only account for  the
correlations between $C_l$ implicitly, giving an effective
distribution for each $\tilde{C}_l$ which contains these correlations.
To obtain explicit information about
the full {\em joint} distribution of all
$\tilde{C}_l$ we must go a step further.

While we wish to relegate further study of the properties of the joint
distribution to future work we will now mention  some preliminary results.

The characteristic function of 
the joint distribution of quadratic
forms such as  Eq.\ (\ref{quadform}) has been studied in
the literature \cite{KendallStuart}. Fourier transforming we can
obtain an integral representation of the joint distribution
(analogously to  Eq.\ (\ref{charfunc}) for the 1-point distributions)
\begin{equation}
p\left(\left\{\tilde{C}_l\right\}\right) =
\frac{1}{(2\pi)^{l_{max}}}\int  \frac{\prod_l \left(e^{-i S_l \tilde{C}_l}dS_l\right)}{\left\vert{\openone -
2i\sum_{l'}S_{l'}{\cal M}^{l'}}\right\vert^{{\frac 12}}}  
\end{equation}

Even though we cannot proceed further in the Fourier inversion or must take
recourse to approximate methods, we can still gain useful information
about the joint statistics of the set of $C_l$.
Similarly to subsection \ref{quad:moments} we can  derive the
cumulants of the variates from the logarithm of the integrand
(disregarding the complex exponential).
It is easy to show that for the covariance between the pseudo--$C_l$ we can write
\begin{equation}
\left\langle{\Delta \tilde{C}_l \Delta \tilde{C}_{l'}}\right\rangle={\rm tr}\,
{{\cal{M}}^{(l)}{\cal{M}}^{(l')}}.
\label{covmat}
\end{equation} 
It turns out that higher order joint moments are computed similarly in terms of traces
of products of ${\cal{M}}^{(l)}$ of various $l$. Note that these
expressions are valid in general, but the special form of
${\cal{M}}^{(l)}$ in the case of azimuthal symmetry will reduce the scaling of
the computational cost for their evaluation.

\subsection{Approximations for the non--symmetric case}
\label{approx}
Returning to the case which we have fully solved, let us perform a
reality check on the assumptions we have made.
There are some important situations where the noise pattern does {\em not}
follow the azimuthal symmetry of the survey geometry. In the case of
the Planck satellite the scanning strategy is approximately centered
on the ecliptic poles, while the Galactic cut is 
tilted through $\approx 60^\circ$
with respect to this. In this case  Eq.\ (\ref{probdist}) becomes an
approximation. 
We found it to be very accurate indeed
to continue using these distributions with the
$\sigma^2_{lm}$ computed for an asymmetric $W_N$, even for a strongly
asymmetric noise pattern. This approximation will be worst in the
least interesting, noise dominated regime at very high $l$.
Note that the 
$\sigma^2_{lm}$ and hence the $\langle \tilde{C}_l \rangle$ 
remain exact (because the $\lambda_{lm}$ are independent of
the azimuth) but the remarks leading to  Eq.\ (\ref{probdist}) are no longer
exactly true. For applications
the final justification comes from the excellent agreement we find
when we check against our Monte Carlo simulations which is illustrated
in Figure \ref{pcldists}. In fact all the applications we present in the
following were computed using strongly asymmetric noise patterns.

\section{Monte--Carlo simulations}
\label{quad:results}

To test our formalism we performed 3328 Monte Carlo (MC) 
simulations for a high resolution CMB satellite, such as MAP or
Planck (resulting in a sky coverage comparable to COBE). We simulated
realizations of the CMB sky in the standard cold 
dark matter model($\Omega_m=1$, $\Omega_bh^2=0.015$,
$H_0=70$km/s/Mpc). From 
these maps we carved out a $\pm 20^\circ$ Galactic cut and contaminated the
remaining area with spatially modulated Gaussian white noise of
maximum root mean squared temperature $124\mu K$ per pixel of characteristic size
3.4 arcminutes. We use
a tilted noise template $W_N=\sqrt{\sin\theta_E}$, where $\theta_E$ is the
ecliptic latitude, as a simple 
model of the noise pattern which
would result from scanning along meridians through the
ecliptic poles. We then Fourier analysed these
maps and stored the resulting $\tilde{C}_l$. 

\begin{figure}[tf]
\centerline{\psfig{file=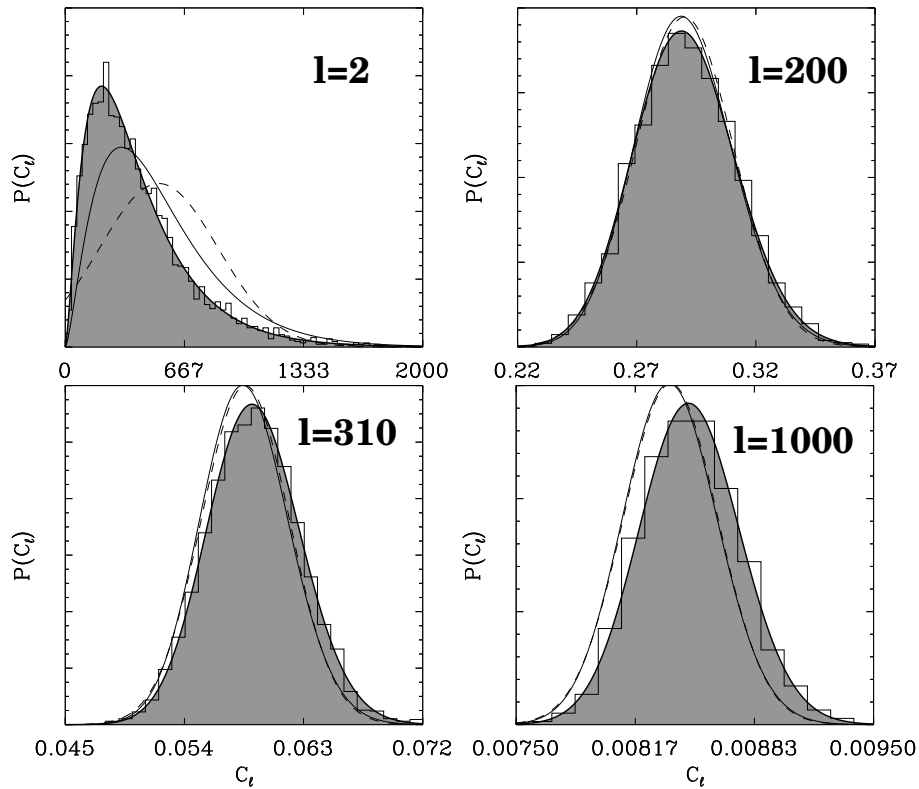,width=.65\textwidth}}
\caption[The pseudo--$C_l$ sampling pdfs. Also shown are the $\chi^2$ and
Gaussian approximations. ]{The pseudo--$C_l$ sampling pdfs , Eq.\
\protect\ref{probdist} (shaded),  
the $\chi^2$ (solid lines) and the Gaussian (dashed) approximations
compared to Monte 
Carlo simulations (histograms) for
l=2,200,310,1000. The theoretical $C_l$ were computed for the standard cold 
dark matter model($\Omega_m=1$, $\Omega_bh^2=0.015$,
$H_0=70$km/s/Mpc). Note that the Monte Carlo simulations were run
with noise pattern which strongly broke the azimuthal symmetry
assumption. We nevertheless find superb agreement with our analytical
predictions. The $\tilde{C}_l$ are in units of $\mu K$. }
\label{pcldists}
\end{figure}

To give a visual impression of
the resulting probability densities we show four cases
in Figure \ref{pcldists}. These cases were chosen to represent different
regimes. The first case, $l=2$ is especially sensitive to effects due to the cut
sky and an important region for the overall normalisation of the power
spectrum. Two cases for intermediate $l$ ($l=200$, $l=310$)  probe cosmologically
interesting scales (on top of the first acoustic peak and in the
first trough of the standard cold dark matter spectrum we use). At
high $l$, $l=1000$, 
there is a significant noise contribution.

Next to our analytical  results we show the histogrammed 
results from our  MC simulations. A Kolmogorov--Smirnov test failed
to detect deviations between the distributions of this
MC population and  Eq.\ (\ref{probdist}) at 99\% confidence, which
validates our semi--analytic expressions. 

Also shown in Figure \ref{pcldists} are the $\chi^2$
distributions which the $C_l$ would follow in the full sky case as
well as the commonly used Gaussian approximation \cite{jungman}.
These are mean adjusted to account for the lost solid angle due
to the Galactic cut. At $l\lesssim 30$ the difference is
striking. For higher $l$ the Gaussian approximation becomes better as
higher moments die away by dint of the Central Limit Theorem, but
there remain visible systematic differences to the true distributions.
In particular, there is a residual shift in the mean and the
approximations tend to be slightly narrower than the histograms for
very high $l\approx 1000$.

\begin{figure}[tf]
\centerline{\psfig{file=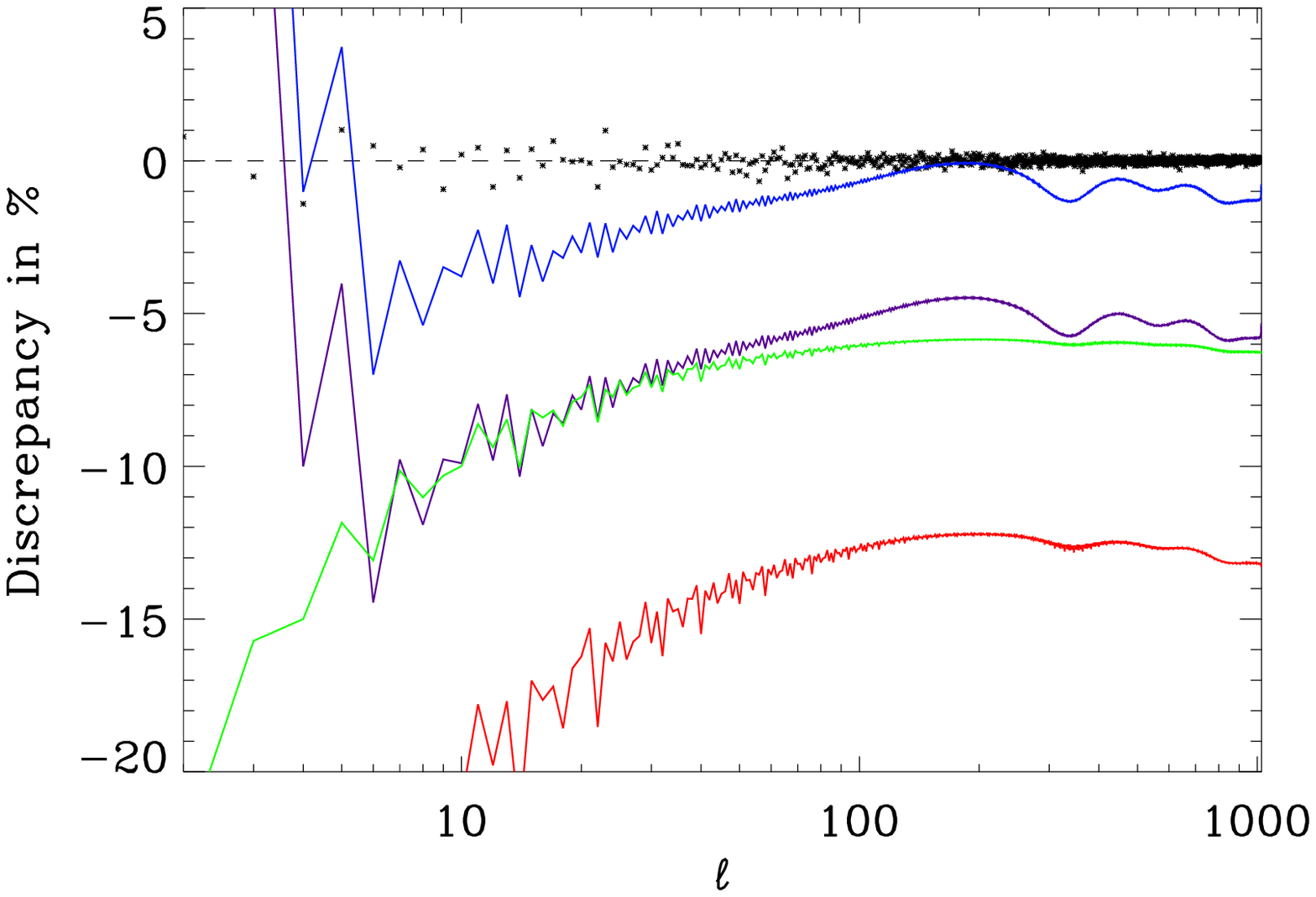,width=.8\textwidth}}
\caption[Percentage discrepancy of the first four moments of the
pseudo--$C_l$ compared to the $\chi^2$ approximation as functions of 
angular wave number $l$.]{Percentage discrepancy of the first four moments of the
pseudo--$C_l$ compared to the $\chi^2$ approximation as a function of 
angular wave number $l$. 
The solid lines are, top to bottom, 
the percentage discrepancy between 
the mean, standard deviation, skewness, and kurtosis of the 
$\chi^2$ approximation
and the pseudo--$C_l$ distributions. The dashed line corresponds to 0\%
discrepancy. The stars are the 
$\langle \tilde{C}_l \rangle$ computed
from 3328 Monte Carlo runs, showing excellent agreement only
limited by Monte Carlo noise to better than 0.1 \% for all $l$.}
\label{moments}
\end{figure}

This becomes a more quantitative observation when looking at
the percentage discrepancies between the mean and variances as a function of
$l$ in Figure \ref{moments}. 
The discrepancy is  of the
order of 1 \% in the mean and 5\% in the standard deviation 
on most scales, except for $l<70$ where the effect is 
larger. These discrepancies are important at the level of
precision of future almost full sky missions. For medium and small
sky coverage the mode couplings are stronger and we expect this to
have an even larger effect on the probability distributions. 
We also compare the 
skewness $\beta_1$ and kurtosis 
$\beta_2$ of the $\chi^2$
distributions to our distributions. The percentage difference is
larger than for the first two moments but arguably less important 
at large $l$, since $\beta_1$ and $\beta_2$ decay as
$(2l+1)^{-\frac{1}{2}}$ and $(2l+1)^{-1}$, respectively. 

\section{Applications}
\label{quad:applications}
\subsection{One--dimensional parameter estimation}
\label{parest}
\begin{figure}[tf]
\centerline{\psfig{file=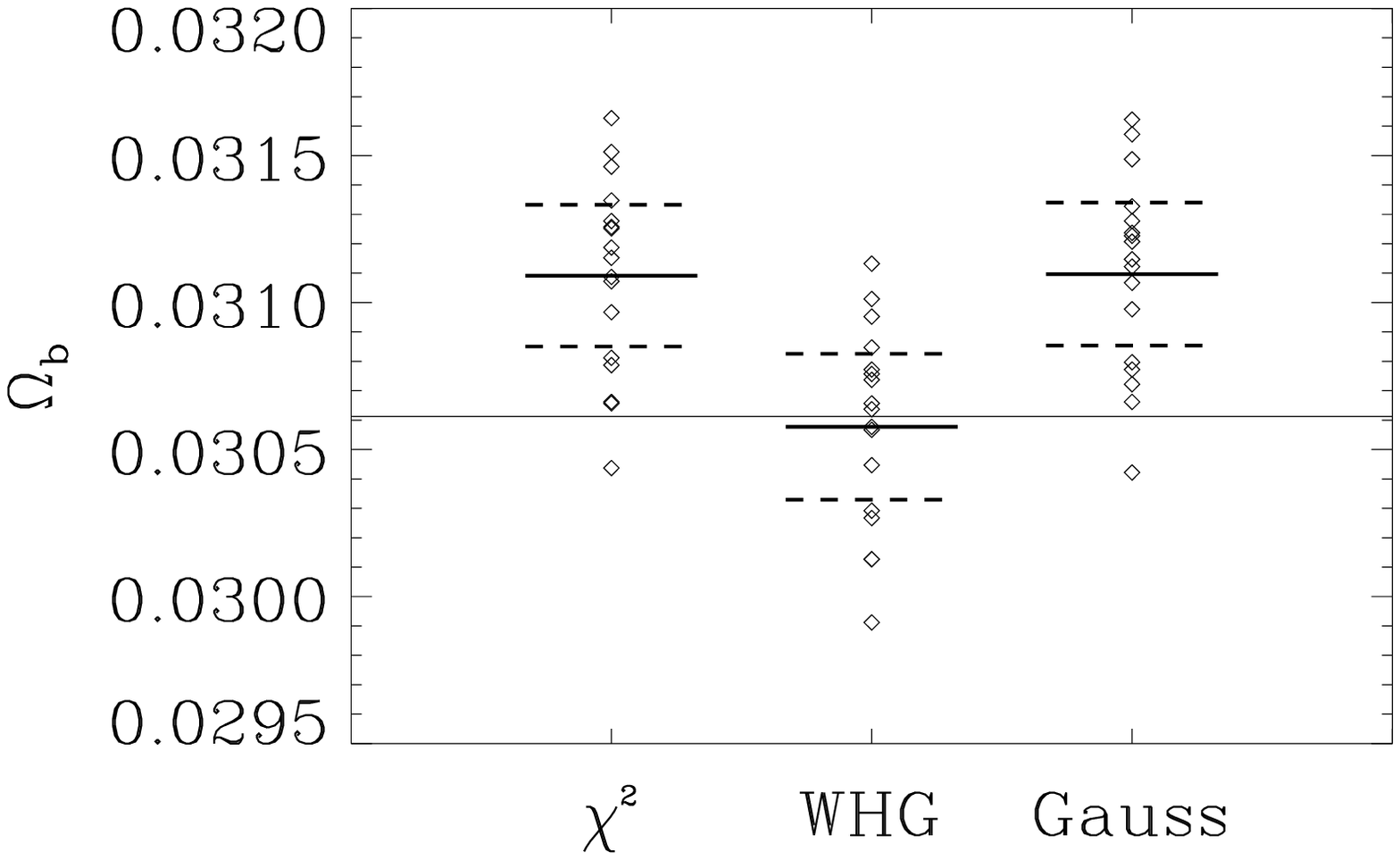,width=.7\textwidth}}
\caption[One parameter likelihood estimates of $\Omega_b$.]
{One parameter likelihood estimates of $\Omega_b$. Shown are the
means (solid) 
within $\pm 3$ standard errors in the mean (dashed) 
of estimations from 17 realizations (diamonds) of
the sky. Both the $\chi^2$ (left)
and the Gaussian (right) approximations produce a bias, 
while our approximation (center) has no detectable bias. }
\label{obbias}
\end{figure}

As a first application we study the effect of
approximating the likelihood for parameter estimation. 
Since our distributions have the correct means, we simply
multiply them together 
for a simple, unbiased approximation to the likelihood
\begin{equation}
\tilde{{\cal L}}= \prod_l p(\tilde{C_l})
\label{likelihood_approx}
\end{equation}
This is a
conservative approximation in the sense that we will not overestimate
the estimation accuracy since the marginal distributions have all correlations
{\em integrated} out and we will therefore overestimate the
error bars on the $C_l$. Using this likelihood, as well as the
Gaussian and $\chi^2$ approximations, we attempt to estimate
the baryon parameter $\Omega_b$ (holding all other parameters
constant) from several randomly selected
realizations in our MC pool. The results are shown in
Figure \ref{obbias}. As expected, Gauss and $\chi^2$  consistently
find estimates which are biased about 1.6\% high, 
3 standard errors of the mean away from the true value, 
while our likelihood gives
a perfect fit. 

Since the moment 
discrepancies depend on the underlying cosmological theory, this level
of bias will also depend on the true theory. Therefore,  this number
should only be taken as indicative of the general level of the error introduced by
using the naive Gaussian or $\chi^2$ approximations.

\subsection{Multi--parameter estimation}
\label{multiparest}

We demonstrated in the previous subsection that we can use our formalism
to construct approximate forms for the likelihood which lead to an
unbiased parameter estimate. The idea proposed in this subsection
is motivated by the fact that higher
moments of the $\tilde{C}_l$ distributions die away quickly with
$l$ ({cf.\ }  Eq.\ (\ref{meanetc})) especially for observation geometries with large
sky coverage. In particular we found in the case we studied that the
distributions were visually indistinguishable from Gaussians for
$l\gtrsim 100$.
For parameters which are sensitive to high $l$ modes, we therefore
further approximate  Eq.\ (\ref{likelihood_approx}) by replacing the
pseudo--$C_l$ pdfs with Gaussians  with the same mean and the same
variance,
\begin{equation}
\widehat{\cal L}(C_l)=\prod_{l>100} \exp\left[-\frac{\left(C_l-\left\langle {\tilde{C_l}} \right\rangle\right)^2}{\left\langle {\Delta \tilde{C}_l^2} \right\rangle}\right]
\label{likelihood_approx_approx}
\end{equation}
In this approximation, maximum likelihood estimation has  reduced to simple $\chi^2$
fitting, however with the correct means and variances which implicitly
account for inter-$C_l$ correlations.

The operation count for each likelihood evaluation has now dropped to
$N_{pix}^{\frac{3}{2}}$. If the computation of the first and second
pseudo-$C_l$ moments is organised in a memory--efficient manner the
elements of the coupling matrix need to be computed only once. We find that we can
evaluate  Eq.\ (\ref{likelihood_approx_approx}) at a rate of several hundred
times per hour. This surprisingly slow scaling of the computational
cost with number of theories is presumably due to more efficient use of
on--processor caching.

\begin{figure}[tf]
\centerline{\psfig{file=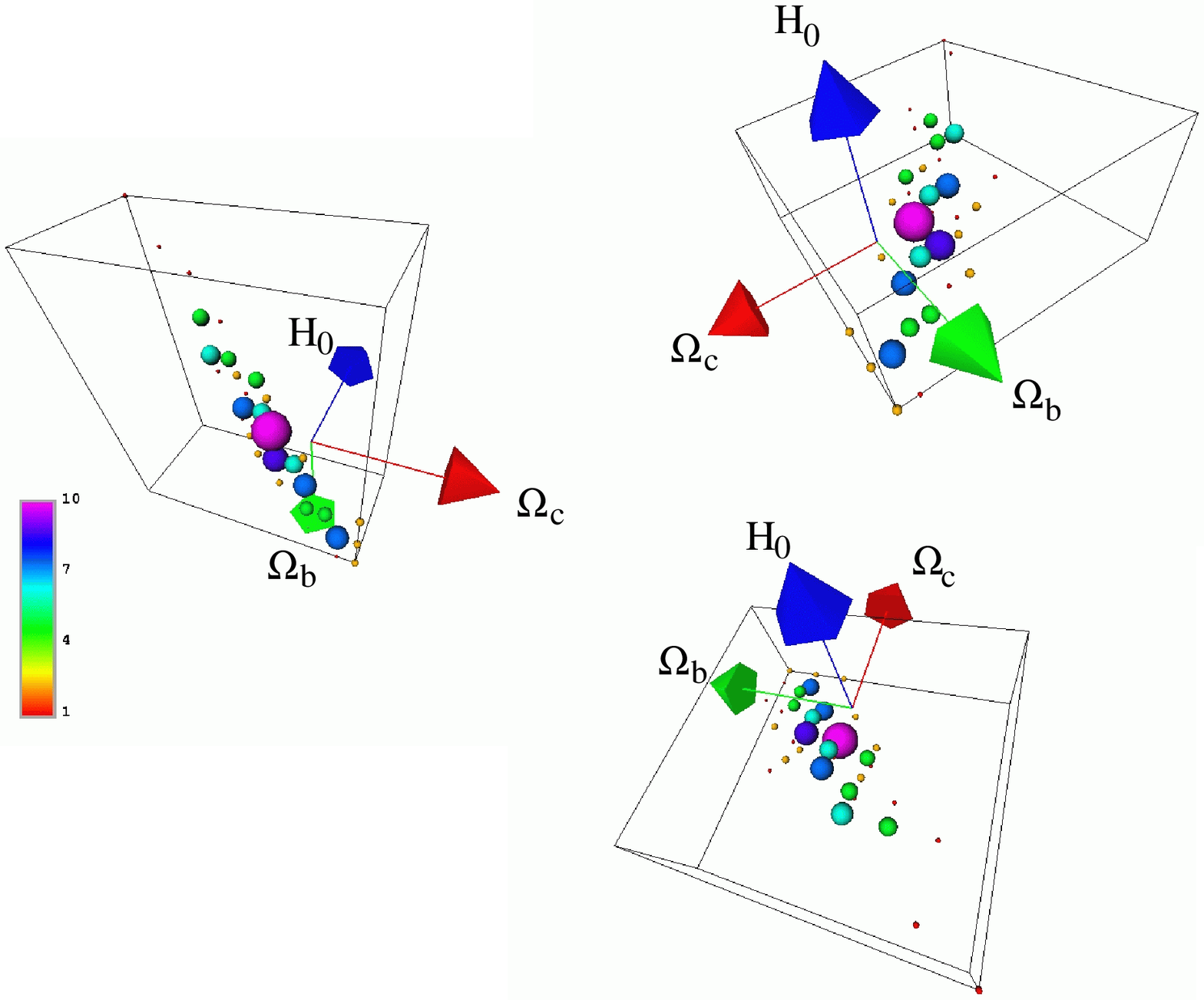,width=\textwidth}}
\caption[Multi--parameter estimation of $\Omega_0$, $\Omega_b$ and $H_0$ using  naive $\chi^2$ fitting]
{Multi--parameter estimation of $\Omega_0$, $\Omega_b$ and $H_0$ using
naive $\chi^2$ fitting. The three panels show 
three views from different directions of the
empirical distribution of the parameter estimates. 
Each sphere represents one bin of the
three--dimensional distribution. The size and colour of a sphere
indicates the number of realisations (out of 100 total) which led to parameter estimates
within its bin. The true
parameter values are at the origin of the coordinate axes. It is clearly
visible that the distribution is shifted with respect to the true
distribution by an amount which is inconsistent with the width of the
distribution. In other words the true values could be ruled out at
high significance if
this estimate was used.}
\label{stupid3D}
\end{figure}

\begin{figure}[tf]
\centerline{\psfig{file=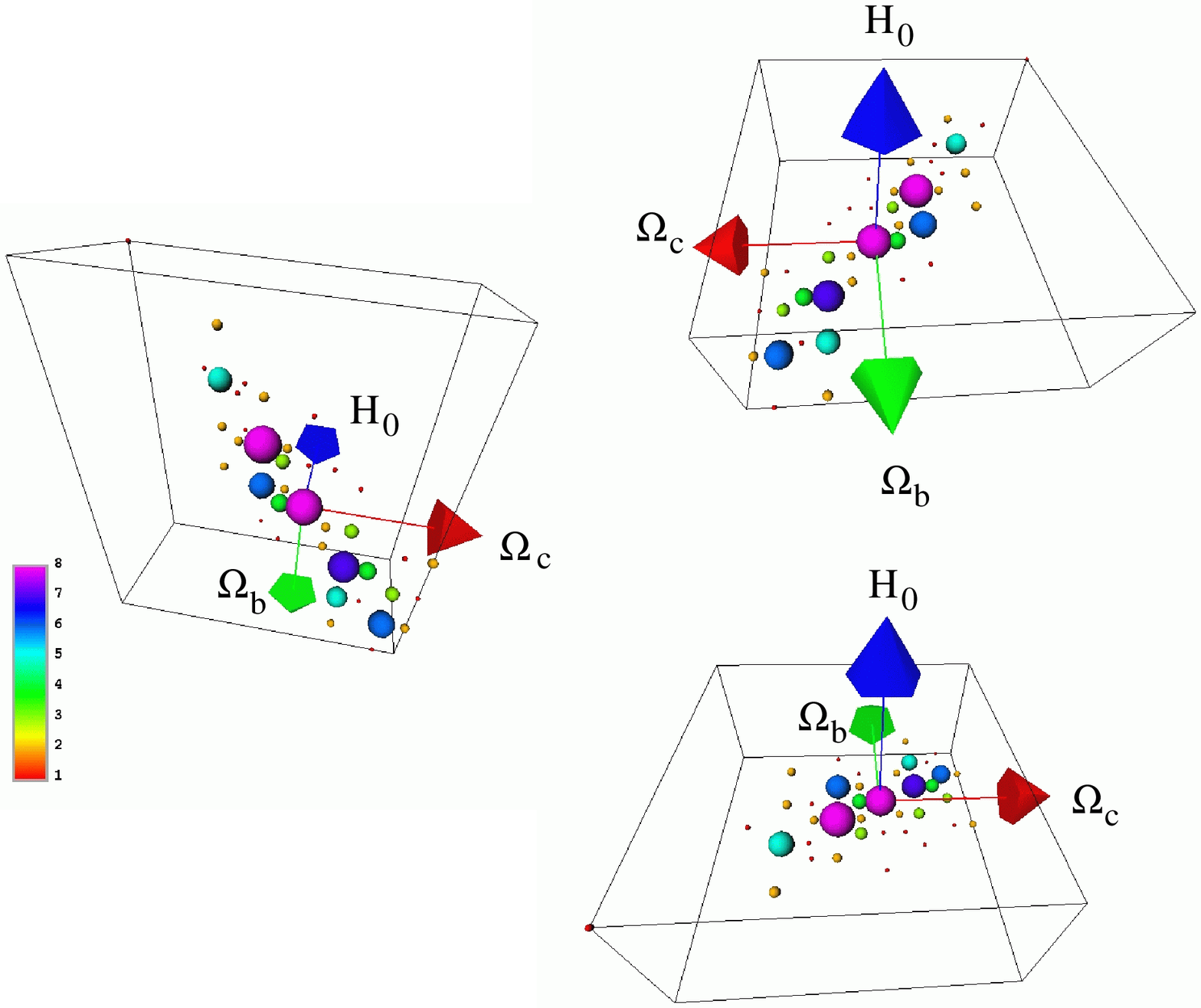,width=\textwidth}}
\caption[Multi--parameter estimation of $\Omega_0$, $\Omega_b$ and $H_0$ using our approximate likelihood.]
{Multi--parameter estimation of $\Omega_0$, $\Omega_b$ and $H_0$ using our approximate likelihood
 Eq.\ (\ref{likelihood_approx_approx}).  The three panels show
three views from different directions of the
empirical distribution of the parameter estimates. 
Each sphere represents one bin of the
three--dimensional distribution. The size and colour of a sphere
indicates the number of realisations (out of 100 total) which led to parameter estimates
within its bin. The true
parameter values are at the origin of the coordinate axes.  It is clearly
visible that the distribution is correctly centered on the true
values. }
\label{pcl3D}
\end{figure}

To illustrate, we solve the problem of estimating 3 parameters
($\Omega_c$ $\Omega_b$ and $H_0$) 
simultaneously from a sky with $12\times 10^6$ pixels. To compare with the naive Gaussian
approach and to show that our method is unbiased, we compute maximum (approximate) 
likelihood estimates from 100
realisations of the sky and plot a representation of the  the
empirical distribution of parameter 
estimates in three dimensions in Figures \ref{stupid3D} (naive $\chi^2$)
and \ref{pcl3D} (our approach).

We find again that our approach is unbiased. The distributions of the
estimates are clearly centered on the true values. 

We stress that our approach avoids the usual difficulties of the
Gaussian approximation as discussed in section \ref{leastsquaresest}. For example, even  
though we use the Gaussian approximation, which  of course does not exclude negative $C_l$,  
they are assigned an exceedingly small probability. This is because no
attempt is made to subtract out 
the noise contribution from the pseudo--$C_l$ --- instead it is
modeled consistently and the (signal$\times$noise) cross term which is present
in each realisation is not allowed to dominate.

\subsection{Test for non--Gaussianity}

We now mention an application which reverses our way of
thinking about the results in this paper and takes advantage of
the exact analytical expressions we compute for the pseudo--$C_l$
distributions. Imagine that we have gained knowledge that a
certain theory is correct, for example through measurements of the
large scale structure of the galaxy distribution and other probes of
cosmology such as supernovae {\em etc}. Then we can use the formalism shown
here as a test of Gaussianity, simply by computing the 
pseudo--$C_l$ and then substituting them into  Eq.\ (\ref{probdist}). An
interesting feature is that this probes Gaussianity scale by scale
which might make it feasible to disentangle confusion effects if the
non--Gaussian signal was only apparent on a small range of scales and
masked by a Gaussian component on other scales. 

\subsection{De--biasing}

In the introduction we mentioned the use of iterative methods to
attack the difficult problem  of  power spectrum estimation. 
A good starting guess can  drastically speed up the  convergence of such methods.
In the case of large sky coverage, where the biases which occur in the
$\tilde{C}_l$ are
small, we suggest the following
recipe as a ``poor man's power spectrum estimator'' for de--biasing
the power spectrum in only ${\cal 
O}\left(N_{pix}^{\frac32}\right)$ computational time: 
\begin{enumerate}
\item compute the $\tilde{C}_l$ from the data, 
\item fit a smooth  curve $C^{smooth}_l$
through them, 
\item compute
${C}^s_l=(\tilde{C}_l-C^{noise}_l)$ and set all negative ${C}^s_l$ to zero,
\item use the ${C}^s_l$ instead of
$C^{theory}_l$ in  Eq.\ (\ref{meanetc}) to compute $\langle\tilde{C}_l\rangle$.
\end{enumerate}
Then  
\begin{equation}
\hat{C}_l=\tilde{C}_l+(C^{smooth}_l-\langle\tilde{C}_l\rangle)
\end{equation}
is an  estimator of the underlying theory $C_l$. There is a tiny
residual  bias which is second order in the discrepancy shown in
Figure \ref{moments}. This is utterly negligible for upcoming satellite
missions. 

\section{Further discussion and conclusions}
\label{quad:conclusions}

We have computed the sampling properties of a set of quantities which
we call pseudo--$C_l$. We have shown the usefulness of these quantities
for the easy compression of large Cosmic Microwave Background data
sets as well as for the forecasting of the discriminatory power of planned
experiments. For example, in situation where percentage accuracy is
required, 
the pseudo--$C_l$ sample variances shown in Figure \ref{moments} could 
immediately replace the usual approximation  which simply
rescales the cosmic variances for the full sky by the sky fraction. 

On the other hand this exact framework can  be used to justify or
derive more approximate methods. We have given an example for the
construction of an approximate but unbiased likelihood for cosmological parameters
on the basis of a Gaussian approximation to the marginalised pseudo--$C_l$ probability
distributions. Using this approximation we can evaluate the likelihood
for hundreds of theories per hour on a single CPU and find unbiased estimates. 

If only a rough treatment of the $C_l$ statistics is sufficient, for
example in the case of large sky coverage,  the results derived
here can be used as a justification for other simplifying assumptions
which are already in use in the literature.

Apart from the obvious applications to experiments with small and
medium sky coverage such as balloons or ground--based missions,
many further uses of this framework are conceivable. For example, one could
use the statistical framework to design 'optimal' scanning strategies,
encoded in $W_N$, and
assess more realistically if secondary 
anisotropies will be detectable with future CMB missions. 
Finally, all ingredients are there to refine the approximation to
the joint likelihood we used in this paper by taking into account the
covariances 
 Eq.\ (\ref{covmat}), for example by using a multivariate Edgeworth expansion
around the peak of the likelihood.

We note that  numerical techniques have recently been developed
\cite{OhSpergelHinshaw} which allow the computational
solution of the power spectrum estimation for high $l$ 
applications
under similar assumptions to the ones which lead to the 
analytical results  derived in this paper.
Their methods are efficient for the case of almost
full sky observations.
A purely numerical approach
still requires significant computational resources, especially if
there is signal in modes with $l>1000$. The results
presented in this paper should
be seen as complementary to such calculations. An analytical framework 
admits a more fundamental approach 
to understanding and is a useful yardstick against which numerical
work can be tested or from which 
approximate methods can be derived as we have demonstrated in the
applications presented in section \ref{quad:applications}.

Our methods do not allow for
significantly correlated noise. Due to the current state of detector
technology, all CMB missions suffer from correlated noise, with the
possible exception of the MAP satellite \cite{Page}. Apart from 
brute force numerical calculation with ${\cal
O}\left(N_{pix}^3\right)$ operations, which is no longer feasible
even with current experiments, there are currently no techniques
for computing sampling statistics in the presence of correlated
noise. This  clearly needs to be addressed as 
an extremely urgent and important issue.

To summarize, we have presented a  theoretical framework for the study
of power spectrum statistics which is applicable 
regardless of the size of the sky area covered.
We go beyond current approximations and present
a semi--analytic formalism for the computation of sampling
distributions of the $C_l$ for any Gaussian cosmological model and a
large and important class of surveying strategies.  
We show that
these results can be usefully applied to the estimation of cosmological parameters
from simulated data.  A range of further applications is suggested which demonstrate the power of
our formalism.

\begin{acknowledgments}

We wish to thank A.\ J.\ Banday  for stimulating discussions. This research
was supported by  Dansk Grundforskningsfonden through its funding for
TAC.

\end{acknowledgments}

\clearpage
\appendix

\section{Low order non--cosmological modes}
\label{quad:loworder}

We assume in this paper that we have access to a partial sky map of
CMB anisotropies free from non--cosmological signal. Even in the event
that foreground contributions can be filtered out using the frequency
information provided by multi--channel experiments, this assumption
merits further discussion --- particularly  for large amplitude low
order modes such as the monopole and the dipole which
are thought to consist exclusively of non-cosmological signal.

An ideal full sky map of the cosmic microwave background
anisotropies has zero mean, by definition. Also, in
Friedmann--Robertson--Walker models, there is no 
cosmological dipole.  But we expect realistic maps resulting  from
observations of the microwave sky to contain both monopole and dipole
components; Galactic emission is positive definite and will therefore
give a monopole contribution. The motion of our solar system with
respect to the frame in which the CMB is at rest produces a
dipole. The question arises how to deal with the presence of such
non--cosmological low order modes. \footnote{We should mention that there is of course a natural occurrence of
spurious monopole and dipole components even in a completely
uncontaminated map as soon as 
a part of the sky is cut away.  This is easy to understand: the
monopole is only constrained to be zero if the anisotropy is
integrated over the full sky. Similarly, a cut sky has a spurious
dipole, even if the dipole was zero on the full sky.
This statistical effect is fully taken into
account in the formalism as it is  presented here.  
All our Monte Carlo simulations of cut skies have   
monopole and dipole components even though they vanish exactly on the full
sky. The sampling distributions of $\tilde{C}_0$ and $\tilde{C}_1$
containing this effect
are quoted in  Eq.\ (\ref{probdist}).}

If the experiment measures total power, we
may hope to remove a sizeable part of the monopole using the frequency
information 
of a multi--channel experiment. In the case of a differential
experiment such as MAP which is blind to the monopole component, the
problem depends on the particulars of the map--making
algorithm employed. Frequency information alone cannot distinguish between a
non-cosmological and a spurious dipole since both are due to the
microwave background. Studies of the peculiar velocity field may
provide some information.

We see the three following different approaches which can be used to deal with the
problem of mode to mode coupling when non--cosmological components are
removed from the sky.

One is orthogonalisation \cite{gorski1}. This solves the problem, but is
not feasible computationally when dealing with millions of degrees of
freedom. 

The other is projecting out the undesired modes either by fitting and
subtracting or more formally by operating with projection
operators. Those approaches lead to a correlation structure for the remaining
modes which is not analytically tractable in the manner presented in
this paper. 

The third is allowing for the presence of a monopole and a dipole in a
Bayesian fashion, by including a rough estimate of their size in the
theory $C_l$. In particular, a rough  {\em a priori} measure of the
monopole and 
dipole present in the data can be input into $C_0$ and $C_1$. Our statistical
framework allows this and yields the correct predicted sampling statistics
for all $C_l$ taking into account all the $l$--$l'$ couplings due to the
cut. In effect, $C_0$ and $C_1$ are treated as nuisance parameters
which are marginalised over in our formalism.

This will change the sampling statistics of low order modes predicted by
our theory in a self-consistent way. Cosmological parameters can then be
estimated in an unbiased fashion without monopole or dipole removal.
Whether this is feasible in practice would be an interesting avenue for
further study. We are not aware of any existing fast method which
offers a similarly consistent treatment.

In practice we observe that for large sky coverage the entries of the
coupling matrices diminish rapidly with increasing $l$. To a good
approximation, modes with 
$l\gtrsim 100$ are therefore unaffected by the presence or absence of
low order modes, $l<2$. One of the main applications of our formalism
has been to the estimation of parameters which depend mostly on the
shape of the $C_l$ spectrum around and beyond the first acoustic
peak. These analyses are unaffected by the details of how low order
modes of non--cosmological origin are dealt with. 

\section{Factorising the correlation matrix}
\label{quad:factorise}

For a general azimuthally symmetric window $W(\theta)$ we can write
the $\tilde{C}_l$ as
\begin{equation}
\begin{split}
\tilde{C}_l=\sum_{m} \int d\Omega_1 W(\theta_1)\int &d\Omega_2 W(\theta_2) Y_{lm}({\bf {\gamma}}_{1}) Y^{\ast}_{lm}({\bf {\gamma}}_{2})\\
&\sum_{l'm'}\sum_{l''m''}a_{l'm'}a^{\ast}_{l''m''}
 Y_{lm}({\bf {\gamma}}_{1})  Y^{\ast}_{l''m''}({\bf {\gamma}}_{2})\nonumber
\end{split}
\end{equation}
We define normalised associated Legendre polynomials 
$\lambda_{lm}(x)$ in  Appendix \ref{quad:couple} and use the notation $\mu=\cos\theta$.
Then  $Y_{lm}(\theta,\phi)=\lambda_{lm}(\mu) e^{im\phi}$ and we notice that the Kronecker--$\delta$
from the azimuthal integration select only one term out of the 
sums over $m'$ and $m''$, namely
\begin{equation}
\tilde{C}_l=\sum_{m} 
\left(\sum_{l'=0}^{\infty}  a^{\ast}_{l'm} {\cal W}_{l'm}^{(l)}\right)
\left(\sum_{l''=0}^{\infty} a^{   }_{l''m} {\cal W}_{l''m}^{(l)}\right)
\end{equation}
where the 
\begin{equation}
{\cal W}_{l'm}^{(l)}\equiv W_{l'm\;lm} =\int 
d\mu W(\mu) \lambda_{lm}(\mu)\lambda_{l'm}(\mu)
\label{eq:Wldef}
\end{equation}
are real matrices. Because the $W$ matrices are real, the terms in the brackets
are complex conjugates of each other and we obtain 
 Eq.\ (\ref{pcldef}) as desired. 

\section{ Algorithm for computing the  coupling matrix}

\label{quad:couple}

Subsection \ref{sphericalstuff}  reviews some useful properties of
spherical harmonics. In subsection \ref{scalprod} we derive an efficient
algorithm for computing the coupling matrix
 Eq.\ (\ref{eq:Wldef}) due to  
an azimuthally symmetric mask,  Eq.\ (\ref{eq:filter}). More general windows can be
piecewise defined in terms of masks of varying height. The
corresponding coupling matrices are then 
suitably weighted  sums over the coupling matrices due to the
individual masks.

%
\subsection{Some properties of spherical harmonics}
\label{sphericalstuff}

The Spherical Harmonics are
\begin{equation}
	Y_{lm}(\theta,\phi) = \lambda_{lm}(\cos\theta) {\rm{ e}}^{{i}
	m\phi}
	\label{eq:ylm_def}
\end{equation}
where the 
\begin{eqnarray}
	\lambda_{lm}(x) &=& \sqrt{ \frac{2l+1}{4\pi}
	\frac{(l-m)!}{(l+m)!} } P_{lm}(x), \quad{\rm for}\,
	m\ge0 	\label{eq:lam_def} \\
	\lambda_{lm} &=& (-1)^m \lambda_{l|m|}, \quad{\rm for}\,
	m\ge0, \nonumber \\
	\lambda_{lm} = 0, \quad{\rm for}\, |m| > l.
\end{eqnarray}
In the following we will only consider positive $m$.
The associated Legendre Polynomials $P_{lm}$ are defined as
\begin{equation}
	P_{lm}(x) = (-1)^m(1-x^2)^{m/2}  \frac{d^m}{dx^m}P_{l\ }(x),
\label{eq:def_leg}
\end{equation}
where $P_{l\ }$ is the Legendre function of the first kind
defined as 
\begin{equation}
	P_{l\ }(x) = \frac{1}{2^ll!}\frac{d^l}{dx^l} (x^2-1)^l.
\end{equation}
They are solution of the differential equation
\begin{equation}
	(1-x^2)\frac{d^2}{dx^2}P_{lm} - 2x \frac{d}{dx}P_{lm}
	+ \left(l(l+1) - \frac{m^2}{1-x^2}\right) P_{lm} = 0
\label{eq:diff_eq}
\end{equation}
Definition  Eq.\ (\ref{eq:def_leg}) leads to a large number of relations
between polynomials of different order $l$ and  $m$ and their
derivatives. Among them the 
following two will be useful (see, {\em e.g.\ } \cite{GradshteynRyzhik} \S 8.700)
\begin{equation}
	P_{lm} = - \frac{x(m-1)}{\sqrt{1-x^2}}P_{l(m-1)} -
	\sqrt{1-x^2}\frac{d}{dx}P_{l(m-1)}
\label{eq:rel_1}
\end{equation}
and
\begin{equation}
	(1-x^2)\frac{d}{dx}P_{lm} = -l x P_{lm} + (l+m) P_{(l-1)m}
\label{eq:rel_2}
\end{equation}

It is possible to obtain the $P_{lm}$ or alternatively the
$\lambda_{lm}$ through relations that raise from the fact that they are
orthogonal polynomials. The following recurrence is stable and very
convenient for numerical uses :

starting with  $\lambda_{00} = 1/\sqrt{4 \pi}$
\begin{eqnarray}
	\lambda_{mm} & = &
	-\sqrt{\frac{2m+1}{2m}}\,\sqrt{1-x^2}\, \lambda_{(m-1) (m-1)}
 	\nonumber \\
	\lambda_{(m+1)m} & = & x\, \sqrt{ 2m + 3}\, \lambda_{mm} \\
	\lambda_{lm}   & = & \left[ x\lambda_{(l-1)m} -
	\frac{ \lambda_{(l-2)m}}{A(l-1,m)} \right]
	A(l,m) \quad{\rm with}\quad A(l,m)= \sqrt{ \frac{4l^2 - 1}{l^2 - m^2} } \nonumber
\end{eqnarray}

%
\subsection{Scalar product on the cut sphere}
\label{scalprod}

The spherical harmonics have the property that they are an orthonormal
basis spanning the Hilbert space of square integrable functions on the
sphere $S^2$, {\em i.e.\ } their scalar product is the identity matrix:
\begin{equation} 
	\int_0^{2\pi} d\phi \int_{-1}^1 d\cos\theta
	Y_{l'm'}(\theta,\phi)Y_{lm}(\theta,\phi) = \delta_{l'l}\delta_{m'm}
\label{eq:scal_prod}
\end{equation}

Let us now define the {\em cut sphere} scalar product of spherical harmonics under a
mask or window $W$  
\begin{equation} 
	A_{l'l}^{mm'}[W]\equiv\int_0^{2\pi} d\phi \int_{-1}^1 d\cos\theta\, W(\theta,\phi)
	Y_{l'm'}(\theta,\phi)Y_{l m}(\theta,\phi) 
\label{eq:scal_prod_cut}
\end{equation}
In the following we will consider only a sharp 'strip' filter whose
boundaries are parallel to the equator (but not necessarily symmetric
about the equator). 
\begin{eqnarray}
	W(\cos\theta) = W(x) &=& 1 \quad{\rm if}\, a \le x \le b
	\nonumber \\
	& = & 0  \quad{\rm otherwise}
\label{eq:filter}
\end{eqnarray}
This is simply related to physically relevant windows. For example
a straight cut removing foregrounds concentrated around the Galactic
disk is modeled by  $\tilde{W} = 1 - W$ with $a=-b$ setting the width
of the cut.
Because $A_{l'l}^{mm'}[\tilde{W}] = {\bf 1} - A_{l'l}^{mm'}[W]$, where ${\bf 1} =
\delta_{l'l}\delta_{m'm}$, everything that follows can be
translated to this window with the replacements
\begin{equation}
	b \le a,\quad {\rm and}\, A_{-1\ }^{-1}[\tilde{W}] = 1.
\end{equation}

Because the window  Eq.\ (\ref{eq:filter}) is independent of $\phi$ it
preserves the azimuthal symmetry and  only spherical
harmonics with the same $m$ are coupled. The non--trivial terms of the
coupling matrix $A$ are then
\begin{eqnarray} 
	A_{l'l}^{m} &=& 2\pi \int_a^b dx\,\lambda_{l'm}(x)\lambda_{lm}(x) \\
	&=&
	\frac{\sqrt{(2l'+1)(2l+1)}}{2}
	\sqrt{\frac{(l'-m)!(l-m)!}{(l'+m)!(l+m)!}}
	\,I_{l'l}^{m}(a,b) 
\end{eqnarray}
where
\begin{equation} 
	I_{l'l}^{m}(a,b) \equiv \int_{a}^{b} dx\, P_{l'm}(x) P_{lm}(x) 
\label{eq:def_coup}
\end{equation}

This expression is symmetric under interchange of  $l'$ and $l$ 
and under $m\rightarrow -m$. If $0\le l\le l_{max}$ this implies that,
asymptotically for large $l_{max}$ there are $\frac{l_{max}^3}{6}$
independent components of 
$I_{l'l}^{m}(a,b)$. If the window $W$ is
symmetric around the equator and 
therefore preserves reflection symmetry, only multipoles with the same
parity of $l$  couple to one another, and the number 
of non--trivial elements reduces to $\sim l_{\rm max}^3/12$.

The problem of evaluating all independent components of
 Eq.\ (\ref{eq:def_coup}) is to find a way of computing 
the integrals  Eq.\ (\ref{eq:def_coup}) without having to take recourse to time--consuming
numerical quadratures.

If one substitutes  Eq.\ (\ref{eq:rel_1}) for $P_{lm}$ and $P_{l'm}$
in  Eq.\ (\ref{eq:def_coup}), integrates by parts to exhibit the second
derivative of $P_{l(m-1)}$, and makes use of the
differential equation  Eq.\ (\ref{eq:diff_eq}) one obtains
\begin{eqnarray}
	I_{l'l}^{m}(a,b) &=&
	-(m+l)(m-l-1)I_{l'l}^{m-1}(a,b) \nonumber \\
	&+& (m-l-1)\left[{xP_{l'(m-1)}P_{l(m-1)}}\right]_{a}^{b} \nonumber \\
	&+& (m+l-1)\left[{ P_{l'(m-1)}P_{(l-1)(m-1)}}\right]_{a}^{b}
\label{eq:rec_int}
\end{eqnarray}
where we used  Eq.\ (\ref{eq:rel_2}) to simplify the last two terms.

This can be simplified further in the case $l\not=l'$, by noting that
the left hand side of  Eq.\ (\ref{eq:rec_int}) is by definition symmetric in
$l',l$ whereas 
the right hand side is not explicitly symmetric.
So, by subtracting  Eq.\ (\ref{eq:rec_int}) from itself after
swapping $l$ and $l'$ one obtains
\begin{eqnarray}
	I_{l'l}^{m}(a,b) &=& \left( (l'-l)\left[{xP_{l'm}P_{lm}}\right]_{a}^{b}
	+ (l +m)\left[{P_{l'  m}P_{(l-1)m}}\right]_{a}^{b}
	- (l'+m)\left[{P_{(l'-1)m}P_{l  m}}\right]_{a}^{b}\right) \nonumber \\
	&/&\left({l'}^2+l'-l^2-l \right) \quad{\rm for}\,\,
	l'\not=l.
\label{eq:int_off_diag}
\end{eqnarray}

The diagonal terms ($I_{l\ }^{m}(a,b) =
\int_a^bP_{lm}P_{lm}$) can be obtained thanks to the
recurrence relation
\begin{equation}
	(l-m)P_{lm} = x(2l-1)P_{(l-1)m} -
	(l+m-1)P_{(l-2)m}.
\label{eq:rec_leg}
\end{equation}
One obtains a recurrence on the diagonal terms involving also the
off--diagonal terms computed by  Eq.\ (\ref{eq:int_off_diag}).
\begin{eqnarray}
	I_{l\ }^{m}(a,b) & = &
	\left(\frac{2l-1}{2l+1}\right)\left(\frac{l+m}{l-m}\right) I_{l-1\ }^{m}(a,b)
	\nonumber \\
	& + &
	\left(\frac{2l-1}{2l+1}\right)\left(\frac{l-m+1}{l-m}\right) I_{(l+1)(l-1)}^{m}(a,b)
	-
	\frac{l+m-1}{l-m} I_{l(l-2)}^{m}(a,b)
\end{eqnarray}

To conclude, the elements of the coupling matrix $A$ are given in
terms of recurrence relations as follows.
For $m\not=m' $ we have
\begin{equation*}
	A_{l'l}^{m'm} = 0 ,
\end{equation*}
while for $ l\not=l'$ 
\begin{equation}
\begin{split}
	A_{l'l}^{m} &=  \Biggl(
		(l'-l)\left[{x\lambda_{l'm}\lambda_{lm}}\right]_{a}^{b}  
		+ \sqrt{ \frac{2l +1}{2l -1} (l ^2-m^2)} 
		\left[{\lambda_{l'm}\lambda_{(l-1)m}}\right]_{a}^{b} - \\
&\quad\quad\quad\quad\quad\sqrt{ \frac{2l'+1}{2l'-1} ({l'}^2-m^2)} 
		\left[{\lambda_{(l'-1)m}\lambda_{lm}}\right]_{a}^{b}
		\Biggr)  \times \frac{2\pi}{(l'-l)(l'+l+1)},
\end{split}
\end{equation}
and all other cases are given by
\begin{equation}
\begin{split}
	A_{l\ }^{m} &= A_{l-1\ }^{m}
	+
	\sqrt{\frac{2l-1}{2l+3}\frac{(l+1)^2-m^2}{l^2-m^2}}
	A_{(l+1)(l-1)}^{m} 
		-
	\sqrt{\frac{2l+1}{2l-3}\frac{(l-1)^2-m^2}{l^2-m^2}}
	A_{l(l-2)}^{m} \quad (l>m)\\
	A_{m\ }^{m} &= A_{m-1\ }^{m-1}
	+ \frac{2\pi}{2m+1}\left[{x\lambda_{mm}\lambda_{mm}}\right]_{a}^{b}
\end{split}
\end{equation}
with $A_{-1\ }^{-1}[W] \equiv 0$. As only a few operations
are needed to evaluate each of these terms, calculating  the whole
coupling matrix costs $O(l_{\rm max}^3)$ operations and only requires about
10 minutes for $l_{\rm max}=2048$ on a fast work--station.

\section{The method of characteristic functions}
\label{quad:cfs}

\subsection{Densities of functions of random variates}
How do we approach computing the probability density function (pdf) of
a (set of) continuous random variate(s) which can itself be written as a function of 
one or more random variates whose pdfs are known? The usual expression
for changing the measure is
\begin{equation}
p(f(x))=p(x)\left|\frac{\partial f}{\partial x}\right|^{-1}
\label{eq:Jacobian}
\end{equation}
where $\left|\frac{\partial f}{\partial x}\right|$ is the Jacobian of the
transformation. However this expression is inconvenient  if the
transform is not monotonic or the 
number of transformed variates is not the same as the number of
original variates. 

In order to attack this more general situation we replace
 Eq.\ (\ref{eq:Jacobian}) by a more powerful
integral expression. The idea here is to
define an "atomic" pdf which allows for only one event. For a continuous
variate this is the delta function.
The sum of these "atoms" over all  possible outcomes, weighted with
the probability that the underlying variates produce this outcome
gives the resulting pdf. In symbols, if the $n$ underlying variates
$X=\{x_i,i=1,\dots,n \}$ have the joint pdf $p(X)$, then the pdf of $z=f(X)$   is
\begin{equation}
p(z)=\int \delta \left(z- f(X)\right) p(X) dX,
\label{fpdf}
\end{equation}
where $dX$ denotes $\prod_{i=1}^{i=n} dx_i$. 

\subsection{The Method of characteristic functions}
In particular we are often interested in the distribution of a linear
combination of $n$ independent  
quantities whose distributions are known. 
Computing these distributions is now straightforward -- we can apply
 Eq.\ (\ref{fpdf}) with the simplification that $f(X)$  
is linear and p(X) factorises owing to the independence of the
variates. The trick is to perform the resulting convolution in Fourier
space by application of 
the Fourier convolution theorem. This is far easier to do than
computing $n$ convolutions, at least if 
$n>2$. The Fourier transform of a pdf of a variate is called its
{\em characteristic function} --- hence the name of the method.
All major pdfs have tabulated Fourier transforms.
This makes this method very simple to apply, up to the final step of
inverse transforming to obtain the required pdf. This may be tabulated
too. If not, extending the integrand to the complex plane and using
contour integration perhaps after 
expanding into partial fractions may lead to success. In any case, at
least only one integral needs to be evaluated
compared to $n$ for the direct convolution.

Even if the inverse transform does not succeed, all is not lost. A key
fact about characteristic functions allows 
us to gain as much information as we like about the statistical
properties of $z$. By formally computing the Maclaurin series of
the Fourier transform which
defines the characteristic function $c(k_z)$ of a variate $z$, we find
that  it
is the generating function for its  statistical moments,
\begin{equation}
c(k_z)=\sum_{n=0}^{n=\infty} \left\langle {z^n} \right\rangle  \frac{(i k_z)^n}{n!}.
\end{equation}
Therefore, 
the moments can be obtained by straightforward differentiation of
$c(k_z)$. Equivalently, and sometimes more conveniently,
the natural logarithm of the characteristic
function generates the cumulants of the distribution. Moments and
cumulants are algebraically simply related.
Once these moments or cumulants are known there is a
host of methods for using them to approximate $p(z)$
\cite{KendallStuart}. In particular, when $n$ is large, the {\em large
number  theorem} states that $p(z)$ approaches a Gaussian with mean
$\left\langle {z} \right\rangle$ and variance $\left\langle {z^2} \right\rangle$  
(under weak conditions on the $p_i(x_i)$). 


\begin{thebibliography}{10}

\bibitem{nonsatellite}
For a compendium of links to experiments refer to {\em e.g.\ } \\ {\tt
  http://www.mpa-garching.mpg.de/$\sim$banday/CMB.html}\\ or\\ {\tt
  http://www.sns.ias.edu/$\sim$max/cmb/experiments.html}.

\bibitem{map}
C.~L. Bennett {\it et~al.}, {\em ``Microwave Anisotropy Probe: A {MIDEX}
  Mission Proposal''}, 1996, see also {\em http://map.gsfc.nasa.gov/}.

\bibitem{planck}
Bersanelli {\it et~al.}, {\em ``{COBRAS/SAMBA}: Report on the Phase {A}
  Study''}, 1996, see also {\em http://astro.estex.esa.nl/Planck/}.

\bibitem{Knox95}
L.~E. Knox, \prd {\bf 52},  4307  (1995).

\bibitem{jungman}
G. Jungman {\it et~al.}, \prd {\bf 54},  1332  (1995).

\bibitem{ZSS}
M. Zaldarriaga, D. Spergel, and U. Seljak, {Astrophys.\ J.\ } {\bf 488},  1
  (1997).

\bibitem{BET}
J.~R. Bond, G. Efstathiou, and M. Tegmark, MNRAS\ {\bf 291},  L33  (1998).

\bibitem{lineweaver1}
C.~H. {Lineweaver} and D. {Barbosa}, A\&A {\bf 329},  799  (1998).

\bibitem{lineweaver2}
C.~H. {Lineweaver} and D. {Barbosa}, \apj {\bf 496},  624  (1998).

\bibitem{SE}
R. {Stompor} and G. {Efstathiou}, MNRAS\ {\bf 302},  735  (1999).

\bibitem{banday}
A.~J. Banday {\it et~al.}, {Astrophys.\ J.\ } {\bf 475},  393  (1997).

\bibitem{gorski0}
K.~M. G\'orski, {Astrophys.\ J.\ Lett.\ } {\bf 430},  L85  (1994).

\bibitem{gorski1}
K.~M. G\'orski {\it et~al.}, {Astrophys.\ J.\ Lett.\ } {\bf 430},  L89  (1994).

\bibitem{gorskimoriond}
K.~M. G\'orski,   (1997), proceedings of the XXXIst Recontres de Moriond,
  "Microwave Background Anisotropies".

\bibitem{BJK}
J.~R. Bond, A.~H. Jaffe, and L.~E. Knox, \prd {\bf 57},  2117  (1998).

\bibitem{borrill}
J. Borrill, Phys. Rev. {\bf D59},  027302  (1999).

\bibitem{tegmark}
M. Tegmark, \prd {\bf 55},  5895  (1997).

\bibitem{BondJaffeKnox2}
J.~R. Bond, A.~H. Jaffe, and L.~E. Knox, 1998, astro-ph/9808264.

\bibitem{Peebles}
P.~J.~E. {Peebles}, {Astrophys.\ J.\ } {\bf 185},  413  (1973).

\bibitem{HauserPeebles}
M.~G. {Hauser} and P.~J.~E. {Peebles}, {Astrophys.\ J.\ } {\bf 185},  757
  (1973).

\bibitem{KendallStuart}
A. Stuart and J.~K. Ord, {\em Advanced Theory of Statistics} (Edward Arnold,
  London, 1994).

\bibitem{GradshteynRyzhik}
I.~S. Gradshteyn and I.~M. Ryzhik, {\em Table of Integrals, Series and Products
  (Fifth Edition)} (Academic Press, London, 1994).

\bibitem{OhSpergelHinshaw}
S.~P. {Oh}, D.~N. {Spergel}, and G. {Hinshaw}, \apj {\bf 510},  551  (1999).

\bibitem{Page}
L. Page, 1998 (private communication).

\end{thebibliography}

\end{document}